\documentclass{amsart}
\usepackage[dvips]{graphics,epsfig}   
\usepackage{graphics}
\usepackage{graphicx,color}
\usepackage{hyperref}
\hypersetup{colorlinks=true,
			linkcolor=magenta,
			citecolor=blue}

\theoremstyle{definition}

\theoremstyle{remark}

\numberwithin{equation}{section}

\begin{document}

\title{Conservative Formulation for Compressible Multiphase Flows}

\author{Evgeniy Romenski}
\address{Sobolev Institute of Mathematics, Novosibirsk, 630090, Russian Federation }
\email{\href{mailto:evrom@math.nsc.ru}{\nolinkurl{evrom@math.nsc.ru}}}
\thanks{The financial support from the Russian Foundation for Basic Research (grants 13-05-00076 and 13-05-12051), Presidium of Russian Academy of Sciences (Programme of Fundamental Research No 15, project 121), and the Siberian Branch of Russian Academy of Sciences (Integration Projects No 127 and No 30) is greatly acknowledged.
}

\author{Alexander A. Belozerov}
\address{Novosibirsk State University, Novosibirsk, 630090, Russian Federation}
\email{\href{mailto:belozerov314@gmail.com}{\nolinkurl{belozerov314@gmail.com}}}

\author{Ilya M. Peshkov}
\address{Sobolev Institute of Mathematics, Novosibirsk, 630090, Russian Federation}
\curraddr{Aix-Marseille Universit\'{e}, CNRS, IUSTI UMR 7343, Marsielle, France}
\email{\href{mailto:peshkov@math.nsc.ru}{\nolinkurl{peshkov@math.nsc.ru}}}
\thanks{The financial support from the Labex MEC (ANR-10-LABX-0092) and A*MIDEX project (ANR-11-IDEX-0001-02), funded by the ``Investissements d'Ave\- nir'' French Government program managed by the French National Research Agency (ANR) is acknowledged}





\begin{abstract}
Derivation of governing equations for multiphase flow on the base of thermodynamically compatible systems theory is presented. The mixture is considered as a continuum in which the multiphase character of the flow is taken into account. The resulting governing equations of the formulated model belong to the class of hyperbolic systems of conservation laws. In order to examine the reliability of the model, the one-dimensional Riemann problem for the four phase flow is studied numerically with the use of the MUSCL-Hancock method in conjunction with the GFORCE flux.
\end{abstract}

\maketitle

\tableofcontents




\section{Introduction}
The development of advanced computational modelling for compressible multiphase flows is of interest
in a number of scientific and engineering disciplines and many industrial applications.
Although the intensive efforts in multiphase flow modelling have undergone in recent years, many basic physical, mathematical, and computational issues are still largely unresolved.
The  classical approach in the development of multiphase models is based
on the assumption that a multiphase flow can be considered as a set of interacting continua
and described as an averaged continuous medium in which the behaviour of each
phase is governed by the conservation laws of mass, momentum and energy, while the interfacial
interaction is taken into account through differential and algebraic source terms in the phase conservation laws \cite{Ishii1975}.

Both past and current research efforts in relation to multiphase
flow modelling mostly concentrate on two-phase computational
models. These include in particular the single-pressure model for two-phase
compressible flows, which is still used as a basic model in
some industrial computer codes. The governing equations
used in the basic single-pressure model are of mixed hyperbolic/elliptic
type thus making the initial-boundary value problem mathematically
ill-posed. Consequently, computations performed with this model on
coarse meshes or using dissipative numerical schemes yield
reasonable solutions, but when the mesh is sufficiently refined or
more accurate numerical methods are used, the solution does not
converge \cite{Stewart1984}. In order to alleviate the ill-posed behavior of the
single-pressure model, various "hyperbolic" modifications have
been proposed \cite{Staedke2005}. These modifications include extra differential
source terms, often referred to as virtual mass, interfacial
pressure and other forces, which are added to phase momentum
balance equations. The resulting system of governing equations
is hyperbolic, but reduction of the system to a symmetric form as
well as writing all equations in a conservative form can not
be achieved, thus making impossible the implementation of the
modified models in two-phase flows that encompass shock and
contact discontinuities.

Another approach is the two-pressure model proposed by Baer and Nunziato \cite{Baer1986}, according to which two
separate media can be handled by two systems of phase conservation
laws coupled with interfacial exchange terms. Intensive efforts have been made in the study of properties of
the Baer-Nunziato model and its modifications and a number of problems of practical interest have been solved with the
use of these models, see for example~\cite{Saurel1999,LeMetayer2005,Zein2010} and references therein. Even though the equations of abovementioned models are hyperbolic, the system of governing equations can not be transformed to a fully conservative form thus leading to difficulties in the case of discontinuous problems as well as in connection with the implementation of modern high accuracy methods.

The generalization of the Baer-Nunziato approach for the modelling of multiphase flow with the number of phases more than two is not clear and only limited number of papers is devoted to this issue, see, for example \cite{Herard2007}.
The model proposed in this paper is a generalization of two-phase compressible Baer-Nunziato type model for the case of three phase mixture. It turns out that the model is hyperbolic but not all of equations can be written in divergent form.
Thus, despite the above research efforts, up to now there is no conventional
form of the model and its governing equations for multiphase compressible flow.
The main challenge in formulation of the multiphase flow models
is associated with the development of a mathematical model that satisfies three important properties:
(a) hyperbolicity (symmetric hyperbolic system in particular); (b) fully conservative form of the governing equations;
(c) compatibility and consistency of the mathematical model with the thermodynamic laws.
These properties provide a mathematical framework for a theory of different initial-boundary value problems
and allow a development of highly accurate numerical methods. So far, there are not available
governing equations for multiphase flows written in a form that satisfies all the three properties.

Here we present another approach in the mathematical and computational modelling of multiphase
flows beyond conventional approaches by means of the theory for thermodynamically
compatible systems of hyperbolic conservation laws \cite{ Romenski2001,Godunov2003}.
Using this theory we derive classes of hyperbolic conservation-form equations
for multiphase compressible flows admitting a straightforward application of advanced high accuracy numerical methods.
In the recent decade (see \cite{Romenski2004,Romenski2007,Romenski2010,Romenski2012,Mattia,Zeidan2011} and references therein) such an approach has been applied to the modelling of two-phase compressible flows including flows with phase transition.
Some high-order numerical methods have also been developed for the one- and two-dimensional single temperature and isentropic models in the abovementioned papers.

The goal of this paper is to formulate a thermodynamically compatible hyperbolic system of governing equations for multiphase compressible flow with arbitrary number of phases and to study its properties. We consider here a single entropy approximation which is applicable for the flow which is not far from the thermal equilibrium.
The general idea of derivation of multiphase flow model is described in \cite{Romenski2012} and here we elaborate it and present the governing equations in terms of phase parameters and closure relations. As an example of application of the proposed approach we consider a one dimensional four-phase model and study numerically some test problems for this model.

The rest of the paper is organized as follows: in Section~\ref{sec:2} and Section~\ref{sec:3} the thermodynamically compatible master system of governing equations is presented for further formulation of multiphase flow equations.  In Section~\ref{sec:4} and Section~\ref{sec:5} the closure relations and governing equations in terms of phase parameters of state are described. In Section~\ref{sec:6} the one-dimensional governing equations for the four phase flow are presented. Finally, in Section~\ref{sec:7} the three Riemann test problems for four phase flows are solved numerically and the results of computations are discussed.

\section{\label{sec:2}Parameters of state for continual description of multiphase medium}
A multiphase mixture can be considered as a continuous medium, which is characterized by the averaged parameters of state such as density, velocity, temperature. Some additional parameters must be introduced if we want to take into account a multiphase character of the flow.

We consider a multiphase compressible flow with $N$ phases. Assume that each phase with number $k$ is characterized by its own parameters: volume fraction $\alpha_k$, mass density $\rho_k$, and  velocity vector $u^k_i (i=1,2,3)$.
The saturation constraint $\alpha_1+\alpha_2+...+\alpha_N=1$ holds.
All above parameters of state are responsible for the mass transfer. What concerns thermal processes, we assume that the mixture is characterized by the mixture entropy $S$ in order to avoid detailed consideration of the heat exchange between phases. Then the mixture temperature will be defined below with the use of laws of thermodynamics. Such an assumption is pure phenomenological and reasonable from the mathematical viewpoint.

Continuum mechanics operates with elements of the medium, characterizing by such parameters as its density, velocity and so on. For the case of multiphase flow we can define mixture density as $\rho=\alpha_1\rho_1+\alpha_2 \rho_2+...+\alpha_N \rho_N $. If to introduce phase
mass concentrations $c_k=\alpha_k \rho_k/\rho$, $(k=1,...,N,~c_1+c_2+...+c_N=1)$, then the average velocity $u_i=c_1 u^1_i+c_2
u^2_i+...+c_N u^N_i$ can be defined.

Additional kinematic parameters characterizing multiphase flow are
relative velocities. Let us choose the velocity of some phase (let it be $N$-th phase) as the basic velocity, then the motion of all other phases can
be characterized by the velocity relative to the chosen one. Thus
we introduce relative velocities as the new multiphase flow
parameters of state: $w_i^1=u_i^1-u_i^N,
w_i^2=u_i^2-u_i^N,...,w_i^{N-1}=u_i^{N-1}-u_i^N.$

Summarizing all above, we conclude that the set of parameters of state which fully describe the multiphase flow as a continuum is:
\[
u_1,u_2,u_3,w^1_1,w^1_2,w^1_3,...,w^{N-1}_1,w^{N-1}_2,w^{N-1}_3,\alpha_1,\alpha_2,...,\alpha_{N-1},
\rho,c_1,c_2...,c_{N-1},S.
\]
All other parameters of state can be derived by these variables with the use of laws of thermodynamics which are formulated below.

\section{\label{sec:3}Generating system of conservation laws for multiphase medium}
In this Section, the master system of hyperbolic conservation laws is formulated, which generates governing equations of multiphase flow. The classical basic governing equations of multiphase continuum are the total mass, total momentum and total energy conservation laws. But the evolution in time of introduced above new state variables should be governed by additional conservation-form equations, the derivation of which is based on paper \cite{Romenski2001}.

\subsection{Master system of conservation laws and its symmetric hyperbolic form}
Consider the fluid flow in Cartesian coordinate system
$x_1,x_2,x_3$. Below the hyperbolic system of conservation-form equations is formulated and its mathematical properties are studied. On this stage we ignore irreversible dissipative processes, and they will be taken into account in the next Section by introduction source terms into some equations. The complete system of governing equations is written in terms of independent flow variables
\[
\alpha_1,\alpha_2,...,\alpha_{N-1},\rho,u_i,c_1,c_2,...,c_{N-1},w^1_i,
w^2_i,,...,w^{N-1}_i,S.
\]
Assume that the generalized specific energy $E$ is defined as a
function of $\alpha_j,c_j,w^j_i ~(j=1,...N-1, i=1,2,3)$, $\rho$
and $S$, then the system which will be used for the derivation of multiphase flow model equations reads as (the summation convention for repeated indices is implied)
\begin{align}
&\frac{\partial \rho S}{\partial{t}} +
\frac{\partial \rho S u_{k}}{\partial{x}_{k}}= 0,
\nonumber \\
&\frac{\partial \rho \alpha_j}{\partial{t}} +
\frac{\partial \rho \alpha_j  u_{k}}{\partial{x}_{k}} = 0, \quad
j=1,...,N-1,
 \nonumber \\
 &\frac{\partial \rho }{\partial{t}} +
\frac{\partial \rho u_{k}}{\partial{x}_{k}}= 0,
\nonumber \\
 &\frac{\partial \rho u_{l} }{\partial{t}}+
\frac{\partial (\rho u_{l} u_{k} + \rho^2 E_\rho \delta_k^l + \rho
w_l^n E_{w^n_k})}{\partial{x}_{k}}= 0, \quad l=1,2,3,
 \label{gensystem} \\
& \frac{\partial \rho c_j }{\partial{t}} + \frac{\partial(\rho
u_{k} c_j + \rho E_{w^k_j})}{\partial{x}_{k}} = 0, \quad
j=1,...,N-1,
\nonumber\\
& \frac{\partial  w^j_k }{\partial{t}}+ \frac{\partial (u_{l}
w_l^j + E_{c_j})}{\partial{x}_{k}} = 0, \quad j=1,...,N-1.
\nonumber
\end{align}
The latter system has two important properties which allow us
to transform it to a symmetric system written in
terms of a generating potential and variables. If the generating
potential is a convex function then the multiphase flow equations
belong to the class of symmetric hyperbolic systems in the sense of Friedrichs \cite{friedrichs1954,DafermosBook}.

The first important property of \eqref{gensystem} is the existence of
compatibility constraints as a steady conservation-form equations
for the vorticity of relative velocities:
\begin{align}
\frac{\partial w_k^j}{\partial x_n}-\frac{\partial w_n^j}{\partial
x_k}=0, \quad j=1,...,N-1; \ k,n=1,2,3. \label{vorticity}
\end{align}
These constraints follow from the equation for the relative velocity.  Actually, if to subtract equation for $w_n^j$ differentiated with respect to $x_k$ from the equation for $w_k^j$ differentiated with respect to $x_n$, we obtain
\[
\frac{\partial}{\partial t}\left( \frac{\partial w_k^j}{\partial
x_n}-\frac{\partial w_n^j}{\partial x_k}\right)=0,
\]
and if the equality \eqref{vorticity} holds for the initial data
$(t=0)$, then it remains valid for~$t>0$.

The second important feature of the above system is that its solution satisfies the first law of thermodynamics, and as a consequence the additional energy conservation law holds. The energy conservation equation can be obtained as a sum of six equations of the system \eqref{gensystem} multiplied respectively by
\[
q_\omega=T=E_S, q_j=E_{\alpha_j} (j=1,...,N-1),
q_0=E-SE_S-VE_V-c_jE_{c_j}-\frac{u_nu_n}{2},
\]
\[
u_l (l=1,2,3), \theta_j=E_{c_j} (j=1,...,N-1), J^j_k=\rho E_{w^j_k} (j=1,...,N-1, k=1,2,3)
\]
and the steady constraint \eqref{vorticity} multiplied by $\rho
u_l E_{w^j_k}$:
\[
\rho u_l E_{w^j_k}\left( \frac{\partial w_k^j}{\partial
x_n}-\frac{\partial w_n^j}{\partial x_k}\right)=0.
\]
Note, that in the definition of $q_0$ the specific volume $V={\rho}^{-1}$ is used for convenience.
As a result, we obtain a conservation-form equation
\begin{align}
&\frac{\partial  \rho (E + {u_{l}u_{l}}/{2})}{\partial{t}} +
\frac{\partial (\rho u^{k}(E + {u_{l}u_{l}}/{2}) + \Pi_k
)}{\partial{x}_{k}} = 0, \label{ENRGeqn}
\end{align}
where $\Pi_k$ is the energy flux vector
\[
\Pi_k=u_k p + \rho u_k  w^l_n E_{w^n_k}+\rho E_{c_j} E_{w^k_j}.
\]
and  $p=\rho^2 E_\rho$.
In order to transform \eqref{gensystem} to a symmetric hyperbolic system, it is necessary to rewrite it in terms of the generating potential and variables. Such a formulation gives us an elegant way to cast equations in a symmetric form. It turns out that the generating potential $L$ can be defined via the total energy $\rho(E+u_lu_l/2)$ by the Legendre transformation. In fact,
as it was noted above, the energy conservation equation can be obtained if to sum equations \eqref{gensystem} and \eqref{vorticity} multiplied by corresponding factors. Thus, the generating potential $L$ can be defined with the use of the following identity
\[
d\rho\left(E+\frac{u_lu_l}{2}\right)=q_\omega d\rho S+q_j d\rho
\alpha_j+ q_0 d\rho + u_l d\rho u_l+ \theta_j d\rho c_j + J^j_k
dw^j_k.
\]
Assuming that $L$ depends on variables $q_\omega,q_j,q_0,u_l,\theta_j,J^j_k$ and denoting
\[
\frac{\partial L}{\partial q_\omega}=\rho S,\frac{\partial
L}{\partial q_j}=\rho \alpha_j,\frac{\partial L}{\partial
q_0}=\rho, \frac{\partial L}{\partial u_l}=\rho u_l,
\frac{\partial L}{\partial \theta_j}= \rho c_j, \frac{\partial
L}{\partial J^j_k}=w^j_k,
\]
we obtain
\[
d\rho\left(E+\frac{u_lu_l}{2}\right)=q_\omega dL_{q_\omega}+q_j
dL_{q_j}+ q_0 dL_{q_0} + u_l dL_{u_l}+ \theta_j dL_{\theta_j} +
J^j_k dL_{J^j_k}=
\]
\[
=d(q_\omega L_{q_\omega}+q_j L_{q_j}+ q_0 L_{q_0} + u_l L_{u_l}+
\theta_j L_{\theta_j} + J^j_k L_{J^j_k}-L).
\]
The latter gives us
\[
L=q_\omega L_{q_\omega}+q_j L_{q_j}+ q_0 L_{q_0} + u_l L_{u_l}+
\theta_j L_{\theta_j} + J^j_k
L_{J^j_k}-\rho\left(E+\frac{u_lu_l}{2}\right)=\rho^2 E_\rho+\rho
w^j_k E_{w^j_k}.
\]

Now all fluxes in equations of system \eqref{gensystem} can be expressed in terms of $q_\omega,q_j,q_0,u_l,\theta_j,J^j_k$ and $L_{q_\omega},L_{q_j},L_{q_0},L_{u_l},L_{\theta_j},L_{J^j_k}$.
Thus, equations \eqref{gensystem} take the following form
\begin{align}
&\frac{\partial L_{q_n}}{\partial t}+\frac{\partial u_k
L_{q_n}}{\partial x_k}=0, \quad n=\omega,1,...,N-1,0, \nonumber \\
&\frac{\partial L_{u^l}}{\partial t}+\frac{\partial (u_k
L_{u^l}+J^j_k L_{J^j_l}-\delta^l_k J^j_n L_{J^j_n})}{\partial
x_k}=0,  \nonumber \\
&\frac{\partial L_{\theta_j}}{\partial t}+\frac{\partial (u_k
L_{\theta_j}+J^j_k)}{\partial x_k}=0, \label{generatingsys} \\
&\frac{\partial L_{J^j_k}}{\partial t}+\frac{\partial (u_n
L_{J^j_n}+\theta_j)}{\partial x_k}=0, \nonumber
\end{align}
and the steady constraints \eqref{vorticity} read as
\begin{align}
\frac{\partial L_{J^j_k}}{\partial x_n}-\frac{\partial
L_{J^j_n}}{\partial x_k}=0. \label{vorticity1}
\end{align}
Finally, the energy conservation law in terms of generating potential and variables takes the form
\begin{align*}
\frac{\partial}{\partial t}\left( q_\omega L_{q_\omega}+q_j
L_{q_j}+ q_0 L_{q_0} + u_l L_{u_l}+ \theta_j L_{\theta_j} + J^j_k
L_{J^j_k}-L\right)+
\end{align*}
\begin{align*}
\frac{\partial}{\partial x_k}\left(u_k (q_\omega L_{q_\omega}+q_j
L_{q_j}+ q_0 L_{q_0} + \theta_j L_{\theta_j} + J^j_k
L_{J^j_k})+J^j_k \theta_j\right)=0.
\end{align*}

Now one can derive an equivalent symmetric form of the system \eqref{generatingsys}. To do this, it is necessary to add
\[
J^j_k \left( \frac{\partial L_{J^j_k}}{\partial
x_n}-\frac{\partial L_{J^j_n}}{\partial x_k}\right)=0
\]
to the second equation of \eqref{generatingsys}, and
\[
u^l\left( \frac{\partial L_{J^j_k}}{\partial x_n}-\frac{\partial
L_{J^j_n}}{\partial x_k}\right)=0
\]
to the last equation of \eqref{generatingsys}.
Then \eqref{generatingsys} can be written as follows
\begin{align*}
&\frac{\partial L_{q_n}}{\partial t}+\frac{\partial (u_k
L)_{q_n}}{\partial x_k}=0, \quad n=\omega,1,...,N-1,0, \nonumber \\
&\frac{\partial L_{u^l}}{\partial t}+\frac{\partial ((u_k
L)_{u^l})}{\partial x_k} +L_{J^j_l}\frac{\partial J^j_k}{\partial
x_k} -L_{J^j_n}\frac{\partial J^j_n }{\partial
x_l}=0,  \nonumber \\
&\frac{\partial L_{\theta_j}}{\partial t}+\frac{\partial (u_k
L)_{\theta_j}}{\partial x_k}+\frac{\partial J^j_k}{\partial x_k}=0, \label{generatingsys} \\
&\frac{\partial L_{J^j_m}}{\partial t}+\frac{\partial (u_k
L)_{J^j_m}}{\partial x_k}+L_{J^j_n}\frac{\partial u_n}{\partial
x_m}-L_{J^j_m}\frac{\partial u_k}{\partial x_k}+\frac{\partial
\theta_m}{\partial x_k}=0. \nonumber
\end{align*}
It easy to see that the quasilinear form of the above system is symmetric. In fact, the two first terms in all equations containing derivatives with respect to $t$ and $x_k$ can be written using a matrices of second derivatives of $L$ and $u_kL$ with respect to the variables $q_\omega,q_j,q_0,u_l,\theta_j,J^j_k$, and these matrices are symmetric. The rest terms are clearly symmetric.

Thus, we have formulated the master system \eqref{gensystem} which
will be used for the design of governing equations of multiphase
compressible flow. All equations of the system are written in a divergent form. The system is hyperbolic if the generating potential $L$ is a convex function of the state variables $q_\omega,q_j,q_0,u_l,\theta_j,J^j_k$. The convexity of $L$ is equivalent to the convexity of the total energy $\rho (E+u_lu_l/2)$ as a function of the variables $\rho S, \rho\alpha_j, \rho, \rho u^l, \rho c_j, w^j_k$, because $L$ and $\rho (E+u_lu_l/2)$ are connected by the Legendre transformation \cite{Godunov2003}.

\subsection{Introduction of source terms into the master system}

We consider only two types of phase interaction -- the phase pressure relaxation to the common value and interfacial friction. The pressure relaxation terms in multiphase compressible flow equations are introduced by analogy with the two-phase flow models \cite{Godunov2003}. They can be introduced as a source terms in the phase volume fraction balance laws. The interfacial friction terms (velocity relaxation or drag force) appear as a source terms in the relative velocity equations. Both relaxation processes lead to the thermodynamically equilibrium state of the flow and their definition must satisfy to the thermodynamic laws. First, the relaxation terms must not affect the total energy conservation law.
Second, the entropy production term in the mixture entropy balance law must be non-negative. We also suppose that the Onsager principle of the symmetry of kinetic coefficients holds.

The velocity relaxation terms violate the steady compatibility condition \eqref{vorticity} and the source terms appear in these equation.
That is why we introduce the relative velocity vorticities as artificial variables (see formula (\ref{artvar}) below). Introduction of such variables save a conservative form of the relative velocity equation. Thus, the extension of the master system for processes with phase interaction reads as:
\begin{eqnarray} \label{gensystem1}
 &&\frac{\partial \rho S}{\partial{t}} +
\frac{\partial \rho S u_{k}}{\partial{x}_{k}}= Q,
\nonumber \\
 &&\frac{\partial \rho \alpha_j}{\partial{t}} +
\frac{\partial \rho u_{k} \alpha_j }{\partial{x}_{k}} = -\Phi_j,
\quad j=1,...,N-1,
 \nonumber \\
 &&\frac{\partial \rho }{\partial{t}} +
\frac{\partial \rho u_{k}}{\partial{x}_{k}}= 0,
\\
 &&\frac{\partial \rho u_{l} }{\partial{t}}+
\frac{\partial (\rho u_{l} u_{k} + \rho^2 E_\rho \delta_k^l + \rho
w_l^n E_{w^n_k})}{\partial{x}_{k}}= 0, \quad l=1,2,3,
 \nonumber \\
&& \frac{\partial \rho c_j }{\partial{t}} + \frac{\partial(\rho
u_{k} c_j + \rho E_{w^k_j})}{\partial{x}_{k}} = 0, \quad
j=1,...,N-1,
\nonumber\\
&& \frac{\partial  w^j_k }{\partial{t}}+ \frac{\partial (u_{l}
w_l^j + E_{c_j})}{\partial{x}_{k}} = -e_{klm}u_l
\omega^j_m-\Lambda^j_k, \quad j=1,...,N-1. \nonumber
\end{eqnarray}
Here, the source terms $-\Phi_j$ and $-\Lambda^j_k$ simulate the
pressure and velocity relaxations respectively, and we define them
as
\begin{equation} \label{sourcetms}
\Phi_j=\rho \sum^{N-1}_{n=1} \phi_{jn}E_{\alpha_n}, \quad
\Lambda^j_k=\sum^{N-1}_{n=1} \lambda^{jn}_kE_{w^n_k}.
\end{equation}
The source term $Q$ in the entropy equation is the mixture entropy production caused by dissipative processes:
\begin{equation*}
Q=\frac{\rho}{E_S}\sum^{N-1}_{j=1}\sum^{N-1}_{n=1}\phi_{jn}E_{\alpha_n}E_{\alpha_n}+
\frac{\rho}{E_S}\sum^{3}_{k=1}
\sum^{N-1}_{j=1}\sum^{N-1}_{n=1}\lambda^{jn}_kE_{w^j_k}E_{w^n_k}.
\end{equation*}
The entropy production must be non-negative, which can be
provided by the positive definiteness of the coefficient matrices
$\phi_{jn}$ and $\lambda^{jn}_1,\lambda^{jn}_2,\lambda^{jn}_3$.

Conditions for the equilibrium state $\Phi_j=0$ and
$\Lambda^j_k=0$ are equivalent to $E_{\alpha_n}=0$ and $E_{w^n_k}=0$
due to the positive definiteness of the matrices of kinetic coefficients.
Below we connect the equation of state $E$ with the equations of state for individual phase, that will give us equilibrium conditions in terms of phase pressures and relative velocities.

The additional source term $-e_{klm}u^l \omega^j_m$ ($e_{klm}$ is the Levi-Civita symbol) appears in
the equation for relative velocities. Here the new artificial
variables $\omega^j_m$ are introduced in order to save
conservative-like form of the equation for the relative velocities. These
variables are the relative velocity vorticities
\begin{equation}
\omega^j_m=e_{mkl}\frac{\partial w^j_k}{\partial x_l},
\label{artvar}
\end{equation}
which are connected by the compatibility equation with the
velocity relaxation source terms:
\begin{equation}
\frac{\partial \omega^j_m}{\partial t}+\frac{\partial(u_l
\omega^j_m-u_k \omega^j_l +e_{mln} \lambda_n^j)}{\partial x_l}=0.
\label{artvareqn}
\end{equation}
The latter can be obtained by the differentiating the relative
velocity equation and using the definition of $\omega^j_m$ \eqref{artvar}.
Note that \eqref{artvareqn} can be rewritten in the form
\begin{align*}
\frac{\partial \omega^j_m}{\partial t}+u_l\frac{\partial
\omega^j_m}{\partial x_l} +\frac{\partial e_{mln}
\lambda_n^j}{\partial x_l}=0,
\end{align*}
because the identity ${\partial \omega^j_m}/{\partial x_m}=0$
holds.

This non-stationary compatibility condition gives us the reason to
treat the term $-e_{klm}u^l \omega^j_m$ in the equation for
relative velocities as a true source term.

One can prove that all introduced  source terms  do not affect the
energy conservation law \eqref{ENRGeqn}, which remains the same for
the system \eqref{gensystem1}.

\section{\label{sec:4}Closure relations}
The master system \eqref{gensystem1} can be used to design a multiphase
flow model as soon as the closure relations are defined. First of all, it is necessary to define
an equation of state in the form of the generalized internal energy $E$ as a function of parameters of state.
Indeed, the energy must be defined by such a way that the obtained governing equations have a specified physical meaning.
Then we can compute all derivatives of the equation of state with respect to the parameters of state which are presented in the governing equations.
After that, only kinetic coefficients must be defined in the relaxation source terms.

Assume the equation of sate for each phase is known in the form of a dependence of the internal energy $e_k$ on the density and entropy, i.e.
$e_k=e_k(\rho_k,S_k), k=1,2,...,N$. Then the pressure and temperature of each phase are computed as
\[
p_k=\rho_k^2\frac{\partial e_k}{\partial \rho_k}, \quad
T_k=\frac{\partial e_k}{\partial S_k}.
\]
The natural way to define an equation of state for the mixture is to take it as an averaged phase specific internal energies and kinematic energy of relative motion:
\begin{equation}
E=c_1e_1+c_2e_2+...+c_Ne_N+
\frac{1}{2}\sum_{j=1}^{N-1}c_jw^j_iw^j_i - \frac{1}{2}\left(\sum_{j=1}^{N-1}c_jw^j_i \right)^2
\label{E}
\end{equation}
It has been noted in Section 2  that in the presented model only the entropy of the mixture is taken as a parameter of state and we should specify the above definition of energy for our needs. Let us consider the velocity independent part of the energy
$ E^0=c_1e_1+c_2e_2+...+c_Ne_N $ and suppose that the mixture is in a thermal equilibrium, that means that phase temperatures are equal ($T_1=T_2=...=T_N=T$).
Then one can define the averaged mixture entropy $S=c_1S_1+c_2S_2+...+c_NS_N $ and assume that the entropy of each phase is the sum of the averaged entropy and its perturbation $S_i=S+\delta S_i$. Note, that $c_1\delta S_1+c_2\delta S_2+...+c_N\delta S_N=0 $ due to definition of $S$.
Thus, we have
\begin{eqnarray} \label{approx}
&&E^0=c_1e_1(\rho_1,S_1)+c_2e_2(\rho_2,S_2)+...+c_Ne_N(\rho_N,S_N)= \nonumber \\
&& c_1e_1(\rho_1,S)+c_2e_2(\rho_2,S)+...+c_Ne_N(\rho_N,S)+ \\
&&c_1\frac{\partial e_1}{\partial S_1}(\rho_1,S)\delta S_1+
c_2\frac{\partial e_2}{\partial S_2}(\rho_1,S)\delta S_2+...
c_N\frac{\partial e_N}{\partial S_N}(\rho_1,S)\delta S_N \nonumber
\end{eqnarray}
Now, using the definition of the phase temperatures, we arrive to $T_i=\frac{\partial e_i}{\partial S_i}(\rho_i,S_i)= \frac{\partial e_i}{\partial S_i}(\rho_i,S)+ \frac{\partial^2 e_i}{\partial S_i^2}(\rho_i,S)\delta S_i$. If to substitute the expression
$$
\frac{\partial e_i}{\partial S_i}(\rho_i,S)=T_i-\frac{\partial^2 e_i}{\partial S_i^2}(\rho_i,S)\delta S_i
$$
into equation \eqref{approx} and assume that $T_i=T+\delta T_i$, we obtain
$$
E^0=c_1e_1(\rho_1,S)+c_2e_2(\rho_2,S)+...+c_Ne_N(\rho_N,S)+o(\delta S_i,\delta T_i).
$$
Thus, if the mixture is not far from the thermal equilibrium then the energy in the form $E^0= c_1e_1(\rho_1,S)+c_2e_2(\rho_2,S)+...+c_Ne_N(\rho_N,S)$ can be used as a component of the generalized internal energy for the mixture with the parameters of state $\rho_1,..., \rho_N,S$. Emphasize that in such a mixture we do not define individual phase temperatures and have only the temperature of the mixture as $T=\frac{\partial E}{\partial S}$.

Definition \eqref{E} of the generalized energy together with the definition of the parameters of
state for the mixture given in Section 2 allow us to express
thermodynamic forces (derivatives of the equation of state) for
the mixture  in terms of thermodynamic forces and parameters of
state for individual phase.

In Section 2, the following physical variables are set as independent
\[
u_1,u_2,u_3,w^1_1,w^1_2,w^1_3,...,w^{N-1}_1,w^{N-1}_2,w^{N-1}_3,\alpha_1,\alpha_2,...,\alpha_{N-1},
\rho,c_1,c_2...,c_{N-1},S.
\]

Now our goal
is to connect thermodynamic forces of the mixture
$E_{w^j_k},E_{\alpha_j},E_{c_j},E_S$ with individual phase
thermodynamic forces $\partial e_j/\partial \rho_j,\partial
e_j/\partial S$ and phase parameters of state. To do this, the
following identities should be used:
\[
d\alpha_N=-d\alpha_1-d\alpha_2-...-d\alpha_{N-1}, \quad
dc_N=-dc_1-dc_2-...-dc_{N-1},
\]
\begin{align}
 d\rho_j=d\left(\frac{\rho
c_j}{\alpha_j}\right)=-\frac{\rho_j}{\alpha_j}d\alpha_j+\frac{c_j}{\alpha_j}d\rho+\frac{\rho}{\alpha_j}dc_j,
\ j=1,2,...,N. \label{dif}
\end{align}
From \eqref{E}, we derive the following identity
\[
dE=(e_1-e_N)dc_1+(e_2-e_N)dc_2+...+(e_{N-1}-e_N)dc_{N-1}+c_1de_1+c_2de_2+...
c_Nde_N+
\]
\[
+c_1(w^1_i-(c_1w^1_i+c_2w^2_i+...+c_{N-1}w^{N-1}_i))dw^1_i+...
\]
\[
+c_{N-1}(w^{N-1}_i-(c_1w^1_i+c_2w^2_i+...+c_{N-1}w^{N-1}_i))dw^{N-1}_i.
\]
Now, using \eqref{dif} we obtain
\[
de_j=\frac{\partial e_j}{\partial \rho_j}d\rho_j+\frac{\partial
e_j}{\partial S}dS =-\frac{\rho_j}{\alpha_j}\frac{\partial
e_j}{\partial \rho_j}d\alpha_j+\frac{c_j}{\alpha_j}\frac{\partial
e_j}{\partial \rho_j}d\rho+\frac{\rho}{\alpha_j}\frac{\partial
e_j}{\partial \rho_j}dc_j+\frac{\partial e_j}{\partial S }dS, \
j=1,...,N-1,
\]
\[
de_N=\frac{\partial e_N}{\partial \rho_N}d\rho_j+\frac{\partial
e_N}{\partial S_N}dS_N=-\frac{\rho_N}{\alpha_N}\frac{\partial
e_N}{\partial \rho_N}d\alpha_N+\frac{c_N}{\alpha_N}\frac{\partial
e_N}{\partial \rho_N}d\rho+\frac{\rho}{\alpha_N}\frac{\partial
e_N}{\partial \rho_N}dc_N+\frac{\partial e_N}{\partial S }dS =
\]
\[
=\frac{\rho_N}{\alpha_N}\frac{\partial e_N}{\partial
\rho_N}(d\alpha_1+...+d\alpha_{N-1})+\frac{c_N}{\alpha_N}\frac{\partial
e_N}{\partial \rho_N}d\rho-\frac{\rho}{\alpha_N}\frac{\partial
e_N}{\partial \rho_N}(dc_1+...+dc_N)+\frac{\partial e_N}{\partial
S }dS.
\]
Note, that for $j=1,2,...,N-1$
\[
w^j_i-(c_1w^1_i+c_2w^2_i+...+c_{N-1}w^{N-1}_i)=(u^j_i-u_i)=\sum_{n=1}^N(u^j_i-u^n_i),
\]
where $u_i=c_1u^1_i+...c_Nu^N_i$.

Finally, we end up with the thermodynamic identity
\[
dE=(e_1+\rho_1\frac{\partial e_1}{\partial \rho_1}
-e_N-\rho_N\frac{\partial e_N}{\partial \rho_N})dc_1
+(e_2+\rho_2\frac{\partial e_2}{\partial \rho_1}
-e_N-\rho_N\frac{\partial e_N}{\partial \rho_N})dc_2+
\]
\[
...+(e_{N-1}+\rho_{N-1}\frac{\partial e_{N-1}}{\partial
\rho_{N-1}}-e_N-\rho_N\frac{\partial e_N}{\partial
\rho_N})dc_{N-1}+
(c_1\frac{\partial e_1}{\partial S}+...+ c_1\frac{\partial e_N}{\partial S})dS+
\]
\[
\frac{1}{\rho}(\rho_N^2\frac{\partial e_{N}}{\partial
\rho_{N}}-\rho_1^2\frac{\partial e_{1}}{\partial
\rho_{1}})d\alpha_1+ \frac{1}{\rho}(\rho_N^2\frac{\partial
e_{N}}{\partial \rho_{N}}-\rho_2^2\frac{\partial e_{2}}{\partial
\rho_{2}})d\alpha_2+...+ \frac{1}{\rho}(\rho_N^2\frac{\partial
e_{N}}{\partial \rho_{N}}-\rho_{N-1}^2\frac{\partial
e_{N-1}}{\partial \rho_{N-1}})d\alpha_{N-1}+
\]
\[
\frac{c_1^2}{\alpha_1}\frac{\partial e_{1}}{\partial
\rho_{1}}+
\frac{c_2^2}{\alpha_2}\frac{\partial e_{2}}{\partial
\rho_{2}}+...
+\frac{c_N^2}{\alpha_N}\frac{\partial e_{N}}{\partial
\rho_{N}}+
\]
\[
+c_1(w^1_i-(c_1w^1_i+c_2w^2_i+...+c_{N-1}w^{N-1}_i))dw^1_i+...
\]
\[
+c_{N-1}(w^{N-1}_i-(c_1w^1_i+c_2w^2_i+...+c_{N-1}w^{N-1}_i))dw^{N-1}_i.
\]
Now, if to take into account that
$\rho_j^2{\partial e_{j}}/{\partial \rho_{j}}=p_j$ and
\[
dE=E_{c_1}dc_1+E_{c_2}dc_2+...+E_{c_{N-1}}dc_{N-1}+
E_{\alpha_1}d\alpha_1+E_{\alpha_2}d\alpha_2+...+E_{\alpha_{N-1}}d\alpha_{N-1}+
\]
\[
+E_\rho
d\rho+E_SdS+E_{w^1_i}dw^1_i+E_{w^2_i}dw^2_i+...+E_{w^{N-1}_i}dw^{N-1}_i,
\]
we conclude that
\begin{eqnarray} \label{derivatives}
&&E_{c_j}=H_j-H_N=(e_j+\frac{p_j}{\rho_j})
-(e_N+\frac{ p_N}{\rho_N}), \ j=1,...,N-1, \nonumber \\
&&E_\rho=\frac{1}{\rho^2}(\alpha_1 p_1+\alpha_2 p_2...+\alpha_N
p_N), \\
&&E_{w^l_i}=c_l\sum_{j=1}^Nc_j(u^l_i-u^j_i), \ i=1,2,3;
\ l=1,2,...,N-1. \nonumber \\
&&E_S=T=c_1\frac{\partial e_1}{\partial S}+...+ c_N\frac{\partial e_N}{\partial S}, \quad E_{\alpha_j}=\frac{1}{\rho} (p_N-p_j), \ j=1,...,N-1. \nonumber
\end{eqnarray}

Thus, in case of single entropy approximation, we have defined a generalized energy $E$ as the sum of mass averaged phase internal energies and kinematic energy of relative motion (4.1). All thermodynamic forces can be expressed in terms of derivatives of $E$ with respect to parameters of state. 
The proof of the convexity of this generalized energy, which is needed for the hyperbolicity, remains an open problem. 
Nevertheless in the numerical implementation of four phase model which is considered in below we observe that the eigenvalues of the one-dimensional system are real in a wide area of parameters of state that gives reason to be sure that the model is hyperbolic.

\section{\label{sec:5}Governing equations in terms of phase parameters}
For the numerical implementation it is more convenient to write the system
of governing equations in terms of phase parameters of state. Using \eqref{derivatives}  one can rewrite equations
\eqref{gensystem1} in terms of individual phase variables and
thermodynamic forces:
\begin{eqnarray}
 &&\frac{\partial \rho S}{\partial{t}} +
\frac{\partial \rho S u_{k}}{\partial{x}_{k}}= Q,
\nonumber \\
 &&\frac{\partial \rho \alpha_j}{\partial{t}} +
\frac{\partial \rho u_{k} \alpha_j}{\partial{x}_{k}} = -\Phi_j,
\quad j=1,...,N-1,
 \nonumber \\
 &&\frac{\partial}{\partial{t}} \left(\sum_{j=1}^N \alpha_j\rho_j\right) +
\frac{\partial}{\partial{x}_{k}} \left(\sum_{j=1}^N \alpha_j\rho_j
u^j_k\right)= 0,
\label{gensystem1a} \\
 &&\frac{\partial}{\partial t} \left(\sum_{j=1}^N \alpha_j\rho_j u^j_l\right)+
\frac{\partial}{\partial{x}_{k}} \left
(\sum_{j=1}^N\alpha_j\rho_ju_l^ju^j_k+
 \delta_k^l\sum_{j=1}^N\alpha_j p_j
\right)= 0, \quad l=1,2,3, \nonumber \\
&& \frac{\partial \alpha_j \rho_j }{\partial{t}} + \frac{\partial
\alpha_j \rho_j u^j_k}{\partial{x}_{k}} = 0, \quad j=1,...,N-1,
\nonumber \\
&& \frac{\partial (u^j_k-u^N_k) }{\partial{t}}+ \frac{\partial
(u^j_lu^j_l/2- u^N_lu^N_l/2 + H_j-H_n)}{\partial{x}_{k}} =
-e_{klm}u^l \omega^j_m-\Lambda^j_k. 
\nonumber
\end{eqnarray}
Here, as it was defined before, $\rho=\sum_{j=1}^N \alpha_j\rho_j$
is the total mixture density, $u_k=\frac{1}{\rho}\sum_{j=1}^N
\alpha_j\rho_j u^j_k$ is the mixture velocity, $p_j=\rho_j^2
{\partial e_j}/{\partial \rho_j}$ is the pressure of $j$-th phase,
$H_j=e_j+p_j/\rho_j $ is the enthalpy of $j$-th
phase, and $T=c_1\frac{\partial e_1}{\partial S}+...+ c_1\frac{\partial e_N}{\partial S}$ is the temperature of the mixture.
The energy conservation law \eqref{ENRGeqn} in terms of phase
parameters of state reads as follows
\begin{equation*}
\frac{\partial}{\partial t}\left(\sum_{j=1}^{N}\alpha_j
\rho_j\left(e_j+\frac{1}{2}u^j_l u^j_l \right) \right) +
\frac{\partial}{\partial x_k}\left(\sum_{j=1}^{N}\alpha_j \rho_j
u_k \left( e_j+\frac{1}{2}u^j_l u^j_l+\frac{p_j}{\rho_j}\right)\right)=0.  \label{ENRGeqn_MOD}
\end{equation*}
The artificial variable $\omega^j_m$ reads as
$\omega^j_m=e_{mkl}{\partial (u^j_k-u^N_k)}/{\partial x_l}$.
Finally, the source terms \eqref{sourceterms} can also be
written in terms of phase pressures and velocities:
\begin{equation*}
\Phi_j= - \sum^{N}_{n=1} \phi_{jn}p_n, \quad
\Lambda^j_k=\sum_{n=1}^{N-1}\lambda^{jn}_k c_n(u^n_k-u_k)=
\sum^{N-1}_{n=1} \lambda^{jn}_k c_n \sum_{l=1}^N
c_l(u^n_k-u^l_k), \label{sourceterms}
\end{equation*}
where $\phi_{jN}=-\sum_{n=1}^{N-1}\phi_{jn}, \ j=1,...,N-1$.

\section{\label{sec:6}Four-phase one-dimensional flow model}
\subsection{Governing equations}
The numerical test problems, presented below, deal with the one-dimensional flow of four phases. The governing equations for the four phase flow can be easily derived from the general multiphase flow equations \eqref{gensystem1a} assuming that $N=4$. Assign $4$-th phase as the basic one, i.e. all relative velocities are counted with the use of the velocity of the $N$-th phase. In the 1D case only one component $u, u^k, k=1,2...,N,$ of the mixture velocity and phase velocity vectors exist. Thus the resulting 1D system reads as follows:
\begin{eqnarray}
 &&\frac{\partial \rho S}{\partial{t}} +
\frac{\partial \rho S u}{\partial{x}}= Q,
\nonumber \\
 &&\frac{\partial \rho \alpha_1}{\partial{t}} +
\frac{\partial \rho u \alpha_1}{\partial{x}} = -\Phi_1,
 \nonumber \\
 &&\frac{\partial \rho \alpha_2}{\partial{t}} +
\frac{\partial \rho u \alpha_2}{\partial{x}} = -\Phi_2,
 \nonumber \\
 &&\frac{\partial \rho \alpha_3}{\partial{t}} +
\frac{\partial \rho u \alpha_3}{\partial{x}} = -\Phi_3,
 \nonumber \\
 &&\frac{\partial\rho}{\partial{t}} +
\frac{\partial \rho u}{\partial{x}}=0,
\nonumber \\
&&\frac{\partial}{\partial t}\sum^{4}_{i=1}\alpha_i\rho_i u^i
+ \frac{\partial}{\partial x}
\sum^{4}_{i=1}\left(\alpha_i\rho_i (u^i)^2+\alpha_i p_i \right)=0,
 \label{gensystem_fourphase} \\
 &&\frac{\partial \alpha_1 \rho_1 }{\partial{t}} + \frac{\partial
\alpha_1 \rho_1 u^1 }{\partial{x}} = 0,
\nonumber\\
&&\frac{\partial \alpha_2 \rho_2 }{\partial{t}} + \frac{\partial
\alpha_2 \rho_2 u^2 }{\partial{x}} = 0,
\nonumber\\
&&\frac{\partial \alpha_3 \rho_3 }{\partial{t}} + \frac{\partial
\alpha_3 \rho_3 u^3 }{\partial{x}} = 0,
\nonumber\\
 &&\frac{\partial (u^1 -u^4) }{\partial{t}}+ \frac{\partial
(u^1u^1 /2- u^4 u^4 /2 + H_1-H_4)}{\partial{x}} =
-\Lambda^1, \nonumber \\
&&\frac{\partial (u^2 -u^4) }{\partial{t}}+ \frac{\partial
(u^2u^2 /2- u^4 u^4 /2 + H_2-H_4)}{\partial{x}} =
-\Lambda^2, \nonumber \\
&&\frac{\partial (u^3 -u^4) }{\partial{t}}+ \frac{\partial
(u^3u^3 /2- u^4 u^4 /2 + H_3-H_4)}{\partial{x}} =
-\Lambda^3, \nonumber
\end{eqnarray}
Here $\rho=\alpha_1\rho_1+\alpha_2\rho_2+\alpha_3\rho_3+\alpha_4\rho_4, p_j=\rho_j^2\partial
e_j/\partial \rho_j, H_j=e_j+p_j/\rho_j$, the source terms are transformed to
\[
\Phi_k= - \sum^{N}_{n=1} \phi_{jn}p_n,
\quad \Lambda^j=
\sum_{n=1}^3 \lambda^{jn} c_n(u^n-u), \quad u=c_1u^1+c_2u^2+c_3u^3+c_4u^4.
\]
It is necessary to emphasize that for the numerical treatment the entropy balance law in the complete system of governing equations must be replaced by the energy conservation law, which reads as follows:
\begin{equation*}
\frac{\partial}{\partial t}\sum^{4}_{i=1}\rho_i\left(e_i+\frac{1}{2}(u^i)^2\right)
+\frac{\partial}{\partial x}\sum^{4}_{i=1}\alpha_i \rho_i u^i\left(e_i+ \frac{(u^i)^2}{2}+\frac{p_i}{\rho_i}\right)=0.
  \label{ENRGeqn2Phase}
\end{equation*}

\subsection{Constitutive relations for the mixture of liquids and perfect gases}
In this Section we describe a set of closure constitutive relations for system \eqref{gensystem_fourphase}. First of all we define the perfect gas equation of state and stiffened gas equation of state. Note that the common way is to use equation of state for gases and liquid as a dependence of internal energy on pressure and temperature. In the present paper equation of state is treated as a dependence of internal energy on density and entropy. These two approaches are equivalent but the latter is more preferable for our model.

We take the perfect gas equation of state in the form
\begin{equation}
e(\rho,S)=A\left(\frac{\rho}{\rho^0}\right)^{\gamma-1} e^{S/c_V},
\label{perfect_gas_EOS}
\end{equation}
where $A=\frac{C^2}{\gamma(\gamma-1)}$, $C$ is the velocity of
sound at normal conditions, $\gamma$ is the adiabatic exponent,
$\rho^0$ is the reference density, $c_V$ is the heat capacity
at constant volume. Then the pressure and temperature are computed
as follows:
\begin{equation*}
p(\rho,S)=\rho^2\frac{\partial e}{\partial \rho}=\rho^0
(\gamma-1)A\left(\frac{\rho}{\rho^0}\right)^\gamma e^{S/c_V}, \quad
 T= \frac{\partial e}{\partial
S}=\frac{A}{c_V}\left(\frac{\rho}{\rho^0}\right)^{\gamma-1}
e^{S/c_V}, \label{P_T}
\end{equation*}
Note that the reference temperature $T^0$ can be defined as
$T^0=A/c_V$.

The stiffened gas equation of state we also define as the dependence of internal energy on the density and entropy
in the form
\begin{equation}
e(\rho,S)=\frac{C^2}{\gamma(\gamma-1)}\left(\frac{\rho}{\rho^0}\right)^{\gamma-1} e^{S/c_V}+\frac{\rho_0 C^2-\gamma p^0}{\gamma \rho}.
\label{stiffened_gas_EOS}
\end{equation}
Then, the pressure and temperature are given by
\begin{equation*}
p(\rho,S)=\rho^2\frac{\partial e}{\partial \rho}=
\frac{\rho^0 C^2}{\gamma}\left(\frac{\rho}{\rho^0}\right)^\gamma e^{S/c_V}-\frac{\rho_0 C^2-\gamma p^0}{\gamma},
\end{equation*}
\begin{equation*}
T= \frac{\partial e}{\partial
S}=\frac{C^2}{c_V\gamma(\gamma-1)}\left(\frac{\rho}{\rho^0}\right)^{\gamma-1}e^{S/c_V}. \label{P_T1}
\end{equation*}
Here $C$ is the velocity of sound at normal conditions, $\gamma$ is the adiabatic exponent,
$\rho^0$ is the reference density, $c_V$ is the heat capacity at constant volume
and $p^0$ is a reference pressure which satisfies the condition $p(\rho^0,0)=0$.

\section{\label{sec:7}Numerical study of the Riemann problem for the four phase flow}

In this Section, we solve numerically some Riemann test problems for the four phase flow in order to study properties of the formulated equations and the physical reliability of the model. 

We begin with the description of a numerical method for solving the presented above one-dimensional system is described.
The system of governing equations under consideration can be written in
the general matrix form of the system of conservation laws
\begin{align}
\frac{\partial {\bf U}}{\partial t}+ \frac{\partial {\bf F} ({\bf
U})}{\partial x}={\bf S} ({\bf U}), \label{symbsys}
\end{align}
where the conservative variable vector reads as
\begin{align*}
{\bf U}=
\left(
\begin{array}{c}
\rho \alpha_1 \\
\rho \alpha_2 \\
\rho \alpha_3 \\
\rho \\
\alpha_1 \rho_1 \\
\alpha_2 \rho_2 \\
\alpha_3 \rho_3 \\
\alpha_1 \rho_1 u^1+\alpha_2 \rho_2 u^2+\alpha_3 \rho_3 u^3+\alpha_4 \rho_4 u^4 \\
u^1-u^4 \\
u^2-u^4 \\
u^3-u^4 \\
\alpha_1 \rho_1 \mathcal{E}_1+\alpha_2 \rho_2  \mathcal{E}_2+\alpha_3 \rho_3  \mathcal{E}_3+\alpha_4 \rho_4  \mathcal{E}_4 \\
\end{array}
\right),
\end{align*}
where $ \mathcal{E}_i=e_i+ \frac{(u^i)^2}{2}+\frac{p_i}{\rho_i}$.
The flux vector reads as
\begin{align*}
{\bf F}({\bf U})= \left(
\begin{array}{c}
\rho u \alpha_1 \\
\rho u \alpha_2 \\
\rho u \alpha_3 \\
\rho u\\
\alpha_1 \rho_1 u^1 \\
\alpha_2 \rho_2 u^2 \\
\alpha_3 \rho_3 u^3 \\
\alpha_1 \rho_1 u^1 u^1 +\alpha_2 \rho_2 u^2 u^2+
\alpha_3 \rho_3 u^3 u^3 +\alpha_4 \rho_4 u^4 u^4
+ \alpha_1 p_1+\alpha_2 p_2+ \alpha_3 p_3+\alpha_4 p_4\\
u^1u^1/2 -u^4u^4/2+H_1-H_4 \\
u^2u^2/2 -u^4u^4/2+H_2-H_4 \\
u^3u^3/2 -u^4u^4/2+H_3-H_4 \\
\alpha_1 \rho_1 u^1\mathcal{E}_1+\alpha_2 \rho_2 u^2 \mathcal{E}_2+\alpha_3 \rho_3 u^3 \mathcal{E}_3+\alpha_4 \rho_4 u^4 \mathcal{E}_4
\\
\end{array}
\right),
\end{align*}
and the source term vector reads as
\begin{align*}
{\bf S}({\bf U})=
\left(
\begin{array}{c}
-\Phi_1 \\
-\Phi_2 \\
-\Phi_3 \\
0 \\
0 \\
0 \\
0 \\
0 \\
- \Lambda_1 \\
- \Lambda_2 \\
- \Lambda_3 \\
0 \\
\end{array}
\right).
\end{align*}

We apply a standard finite-volume method for numerical
approximation of the system \eqref{symbsys}. For the control
volume with dimensions $\Delta t=t^{n+1}-t^n$, $\Delta
x=x_{i+1/2}-x_{i-1/2}$ the
difference equation reads as follows:
\begin{align}
{\bf U}^{n+1}_i={\bf U}^{n}_i-\frac{\Delta t}{\Delta x}({\bf
F}_{i+1/2,j}-{\bf F}_{i-1/2,j})+\int^{t^{n+1}}_{t_n}{\bf S({\bf
U})}dt, \label{FVoL}
\end{align}
where ${\bf U}^{n}_i$ is an approximation to the cell average at
the time moment $t^n$ and ${\bf F}_{i+1/2,j}$
is the numerical fluxes on the corresponding cell interface.
Each difference scheme must be specified by the method of
computing fluxes and source terms in \eqref{FVoL}. Many numerical
methods are based on the flux evaluation obtained as a solution of
the Riemann problem \cite{ToroBook}. The solution of the Riemann
problem can be obtained for the hyperbolic system of conservation
laws with known eigenstructure. In our case of the complex system
of governing equations of multiphase flow the eigenvalues and
eigenvectors can be obtained explicitly only in the case of
reduced isentropic model \cite{ Romenski2004}. In the case of more general
model even eigenvalues of the linearized system can not be found
explicitly. That is why we have implemented a MUSCL method
in conjunction with the GFORCE method \cite{ToroBook}, which is based on the centred difference scheme for flux evaluation.

\subsection{GFORCE flux}

In what follows, the GFORCE flux in conjunction with the MUSCL  method \cite{ToroBook} is described for the one-dimensional system of
conservation laws. In multidimensional problems, the developed
numerical method can be applied for each spatial direction
separately. Consider the following system of conservation laws
\begin{align*}
\frac{\partial {\bf U}}{\partial t}+ \frac{\partial {\bf F} ({\bf
U})}{\partial x}=0,
\end{align*}
and corresponding finite-volume approximation
\begin{align*}
{\bf U}^{n+1}_i={\bf U}^{n}_i-\frac{\Delta t}{\Delta x}({\bf
F}_{i+1/2 }-{\bf F}_{i-1/2 })
\end{align*}

At the time moment $t_n$ and at the cell interface $x_{i+1/2}$ the flux
${\bf F}_{i+1/2}$ can be evaluated as a solution to the Riemann
problem with initial data $U_L, U_R$ which are obtained by the
MUSCL-Hancock method in conjunction with the slope limiter \cite{ToroBook}.
In the test problems presented below we use the {\em minmod} limiter.
Thus a computation of intercell boundary extrapolated values ${\bf U^L_i}, {\bf U^R_i}$ is performed
by the following formulae:
$$
{\bf U^L_i}={\bf \hat U^L_i}-\frac{1}{2}\frac{\Delta t}{\Delta x}
({\bf F}({\bf \hat U^R_i})-{\bf F}({\bf \hat U^L_i})),
$$
$$
{\bf U^R_i}={\bf \hat U^R_i}-\frac{1}{2}\frac{\Delta t}{\Delta x}
({\bf F}({\bf \hat U^R_i})-{\bf F}({\bf \hat U^L_i})),
$$
where
$$
{\bf \hat U^L_i}={\bf \hat U^n_i}-\frac{1}{2} \Delta_i, \quad
{\bf \hat U^R_i}={\bf \hat U^n_i}+\frac{1}{2} \Delta_i,
$$
and $\Delta_i$ is the {\em minmod} limiter slope computed as
$$
\Delta_i=max(0,min(\Delta_{i-1/2},\Delta_{i+1/2})), \quad \Delta_{i+1/2} \ge 0,
$$
$$
\Delta_i=min(0,max(\Delta_{i-1/2},\Delta_{i+1/2})), \quad \Delta_{i+1/2} \le 0,
$$
$$
\Delta_{i-1/2}={\bf \hat U^n_i}-{\bf \hat U^n_{i-1}}, \quad
\Delta_{i+1/2}={\bf \hat U^n_{i+1}}-{\bf \hat U^n_{i}}
$$

After that we
apply the GFORCE flux to the known conservative variables.
The GFORCE flux ${\bf F}_{i+1/2}^{GF}$ is a convex average of
well-known Lax-Friedrichs ${\bf F}_{i+1/2}^{LF}$ and Lax-Wendroff
${\bf F}_{i+1/2}^{LW}$ fluxes the definition of which can be found
in \cite{ToroBook}:
\begin{align} {\bf F}_{i+1/2}^{GF}={\omega \bf
F}_{i+1/2}^{LW}+(1-\omega){\bf F}_{i+1/2}^{LF}, \label{gforce}
\end{align}
where \begin{align} {\bf F}_{i+1/2}^{LF}=\frac{1}{2}({\bf F}({\bf
U}_L)+{\bf F}({\bf U}_R))-\frac{1}{2}\frac{\Delta x}{\delta
t}({\bf U}_R-{\bf U}_L) \label{LF}
\end{align}
and \begin{align} {\bf F}_{i+1/2}^{LW}= {\bf F}({\bf
U}_{LW}),\quad {\bf U}_{LW}= \frac{1}{2}({\bf U}_L+{\bf
U}_R)-\frac{1}{2}\frac{\delta t}{\Delta x}({\bf F}({\bf U}_R)-{\bf
F}({\bf U}_L)). \label{LW}
\end{align}

Here the local time step $\delta t$ is used in the definition of
${\bf F}_{i+1/2}^{LF}$ and ${\bf F}_{i+1/2}^{LW}$. It can be
estimated from the local initial data ${\bf U}_L,{\bf U}_R$ as
$\delta t=K \Delta x/S_{max}$, where $S_{max}$ is the speed of the
fastest wave in the local initial data and $K$ is the local
Courant number ($K=0.9$ is usually taken). Such a choice of
$\delta t$ allows us to remove a dependence of the truncation
error on the reciprocal of the Courant number of difference scheme
and eliminate a peculiar to the centred methods diffusivity. The
coefficient $\omega$ in \eqref{gforce} is taken as
$\omega=\frac{1}{1+K}, \quad 0<K \le 1$.

In \cite{ToroTitarev}, it is reported that the GFORCE flux is
upwind due to the nonlinear dependence of the weight $\omega$ on
the local wave speed, and moreover for the linear advection
equation with constant coefficient the GFORCE flux reproduces the
Godunov upwind flux.

\subsection{Source terms numerical implementation}
System \eqref{symbsys} includes six source terms $\Phi_1,\Phi_2,\Phi_3$, $\Lambda_1, \Lambda_2,\Lambda_3$. The terms
$\Phi_i=\phi_{ij}(p_i-p_j)$ simulate the rate of phase pressures
relaxation. The source terms $\Lambda_1=\lambda_{1} c_1 c_2(u^1_1-u^2_1)$ are responsible for the interfacial friction.

The pressures relaxation rate coefficients $\phi_{ij}$ and interfacial friction coefficient $\lambda_ij$ can be quite big,
therefore the corresponding equations can be stiff. That is why it is reasonable to
use the backward Euler method for time integration of equations
for the volume fraction and relative velocities that leads to the implicit sheme.

First, consider the balance equations for volume fractions
\[
\frac{\partial \rho \alpha_k}{\partial{t}} + \frac{\partial \rho
u \alpha_k}{\partial{x}}= \phi_{k1}p_1+\phi_{k2}p_2+\phi_{k3}p_3+\phi_{k4}p_4, \quad k=1,2,3.
\]
The finite-volume difference approximation for this equations with
the use of backward Euler method reads as
\begin{eqnarray}  \label{VFeqns}
&&(\rho \alpha_k)^{(n+1)}_{i}=(\rho \alpha_k)^{(n)}_{i}-
\frac{\Delta t}{\Delta x}({(\rho \alpha)}_{i+1/2}-{(\rho
\alpha)}_{i-1/2})+ \nonumber \\
&& {\Delta
t}\left(
\phi_{k1}(p_1)_{i}^{(n+1)}+\phi_{k2}(p_2)_{i}^{(n+1)}+\phi_{k3}(p_3)_{i}^{(n+1)}+\phi_{k4}(p_4)_{i}^{(n+1)}
\right)
\end{eqnarray}
where ${(\rho \alpha)}_{i+1/2}$,${(\rho
\alpha)}_{i-1/2}$ are computed through the flux evaluation at the cell interfaces
and $(p_k)_{i}^{(n+1)}$ is the function of $\rho_k,S$.

The simplest implementation of implicit  sceme
which gives a good results in many cases is to assume that $(p_k)_{i}^{(n+1)}=p_k\left(\rho_{i}^{(n+1)},S_{i}^{(n)}\right)$.
Assume that the values
${(\alpha_k\rho_k)}^{(n+1)}_{i}, (k=1,2,3,4)$
are already known from the numerical integration of phase mass conservation laws
and take into account that
\[
(\rho_k)_{i}^{(n+1)}=\frac{{(\alpha_k\rho_k)}^{(n+1)}_{i}}{{(\alpha_k)}^{(n+1)}_{i}}, \ (k=1,2,3),
\]
\[
(\rho_4)_{i}^{(n+1)}=\frac{{(\alpha_4\rho_4)}^{(n+1)}_{i}}{1-(\alpha_1)^{(n+1)}_{i}-(\alpha_2)^{(n+1)}_{i}-(\alpha_3)^{(n+1)}_{i}}.
\]
Then, \eqref{VFeqns} can be treated as a nonlinear system of
algebraic equations for $(\alpha_k)_{j}^{(n+1)}$ and can be solved
by iterative methods.

Note that if the pressure
relaxation is assumed to be instantaneous, then it is necessary to solve the following
system of algebraic equations for $(\alpha_k)_{i}^{(n+1)}$ at each mesh
cell:
\[
p_m((\rho_1)_{i}^{(n+1)})-p_n((\rho_2)_{i}^{(n+1)})=0.
\]

A similar algorithm can be implemented for the relative velocity
relaxation. The finite-volume approximation of balance laws for
the relative velocities in case of constant interfacial friction coefficients $\lambda^{kj}$ reads as ($k=1,2,3$)
\[
(u^k-u^4)^{(n+1)}_i=(u^k-u^4)^{(n)}_i-
\frac{\Delta t}{\Delta x}({(u^ku^k-u^4u^4+H_k-H_4)}_{i+1/2}-
\]
\[
-{(u^ku^k-u^4u^4+H_k-H_4)}_{i-1/2})
- {\Delta t}\sum_{j=1}^3 \lambda^{kj} (c_j(u^j-u))_i^{(n+1)}
\]
From the above difference equations using known  conserved variables on the $(n+1)$ time step, the velocities $(u^k)^{(n+1)}_i$ can be easily computed.

\subsection{Numerical solution of the Riemann test problems}
In this Section, results of the numerical solution of the one-dimensional Riemann test problems are presented.

\begin{figure}
\includegraphics*[trim = 25mm 165mm 0mm 0mm, clip, scale = 0.5]{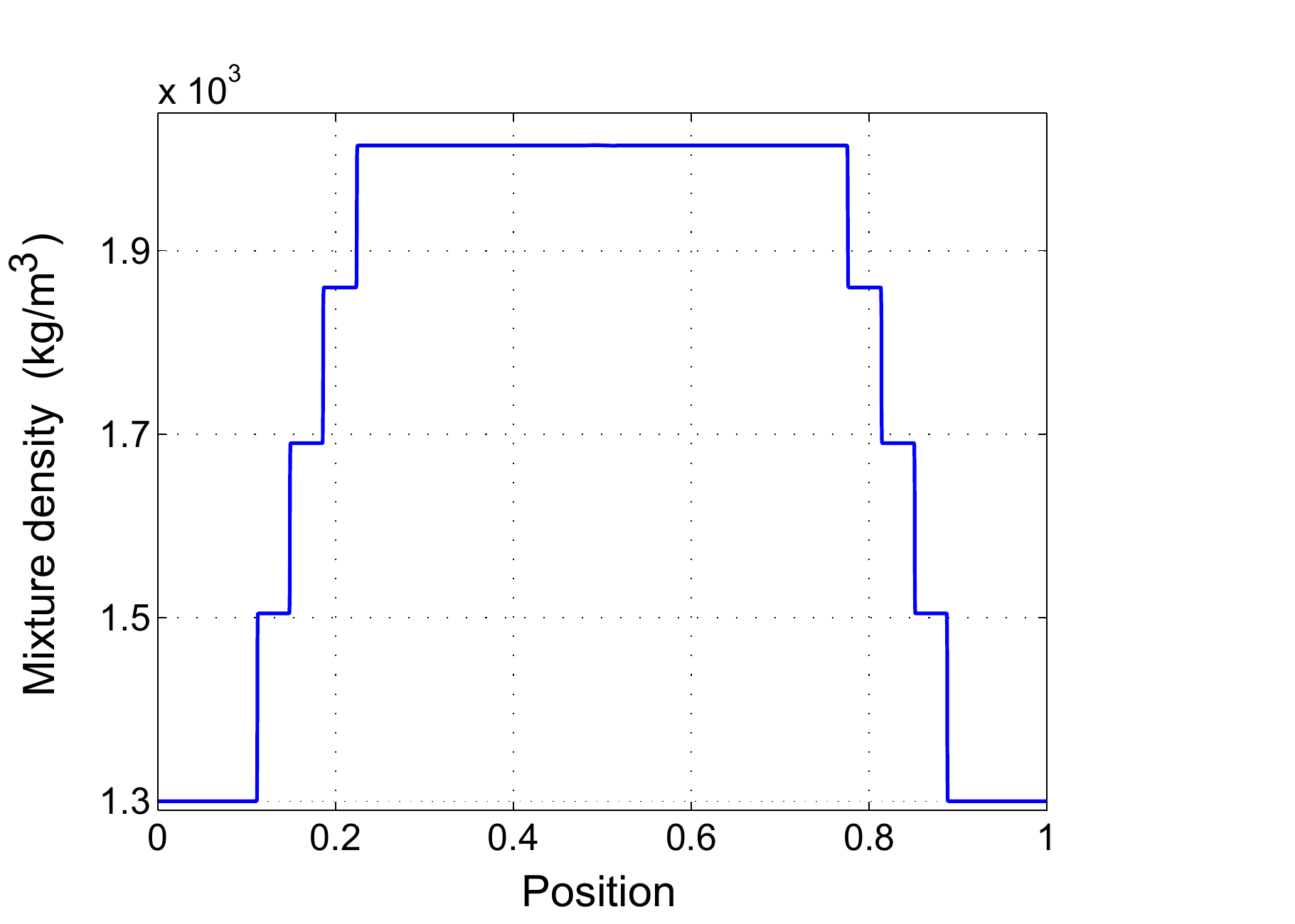}
\caption{Density profile for collision of the mixture of fictitious liquids at the velocity $1000m/s$}
\label{fig:collision}
\end{figure}

The collision and cavitation test problems have been solved for the sake of validation of numerical method and for the study of characteristic properties of the model.
First we consider a symmetric collision of the mixture of four fictitious liquids with the stiffened gas equation of state for liquid \eqref{stiffened_gas_EOS}. The parameters of the equation of state for each phase were taken as
$\rho_{01}=1000kg/m^3$, $\rho_{02}=1200kg/m^3$, $\rho_{03}=1400kg/m^3$, $\rho_{04}=1600kg/m^3$, $C_{01}=1500m/s$, $C_{02}=1700m/s$, $C_{03}=1900m/s$, $C_{04}=2100m/s$, $\gamma_1=\gamma_2=\gamma_3=\gamma_4=2.8$.
Our goal is to demonstrate that the number of wave propagating to both sides of the initial discontinuity coincides with the number of sound waves which is equal to 4 in our case of four phase mixture.
We neglect pressure relaxation and interfacial friction and consider isentropic model assuming $S=0$ in computations. The phase volume fractions are $\alpha_i=0.25, i=1,2,3,4$ and the pressure is atmospheric ($10^5Pa$) in the initial data everywhere. The velocity of collision is $U=1000 m/s$, that is $u_1=u_2= u_3= u_4=U$ on the left side of initial discontinuity and $u_1=u_2= u_3= u_4=-U$ on the right side. Computations were made for 3000 mesh cells, and the mixture density profile is presented on Figure~\ref{fig:collision} at some instant of time. Four left propagating and four right propagating shock waves are clearly seen, which is in accordance with characteristic properties of the governing differential equations.

\begin{figure}
\includegraphics*[trim = 23mm 165mm 0mm 0mm, clip, scale = 0.5]{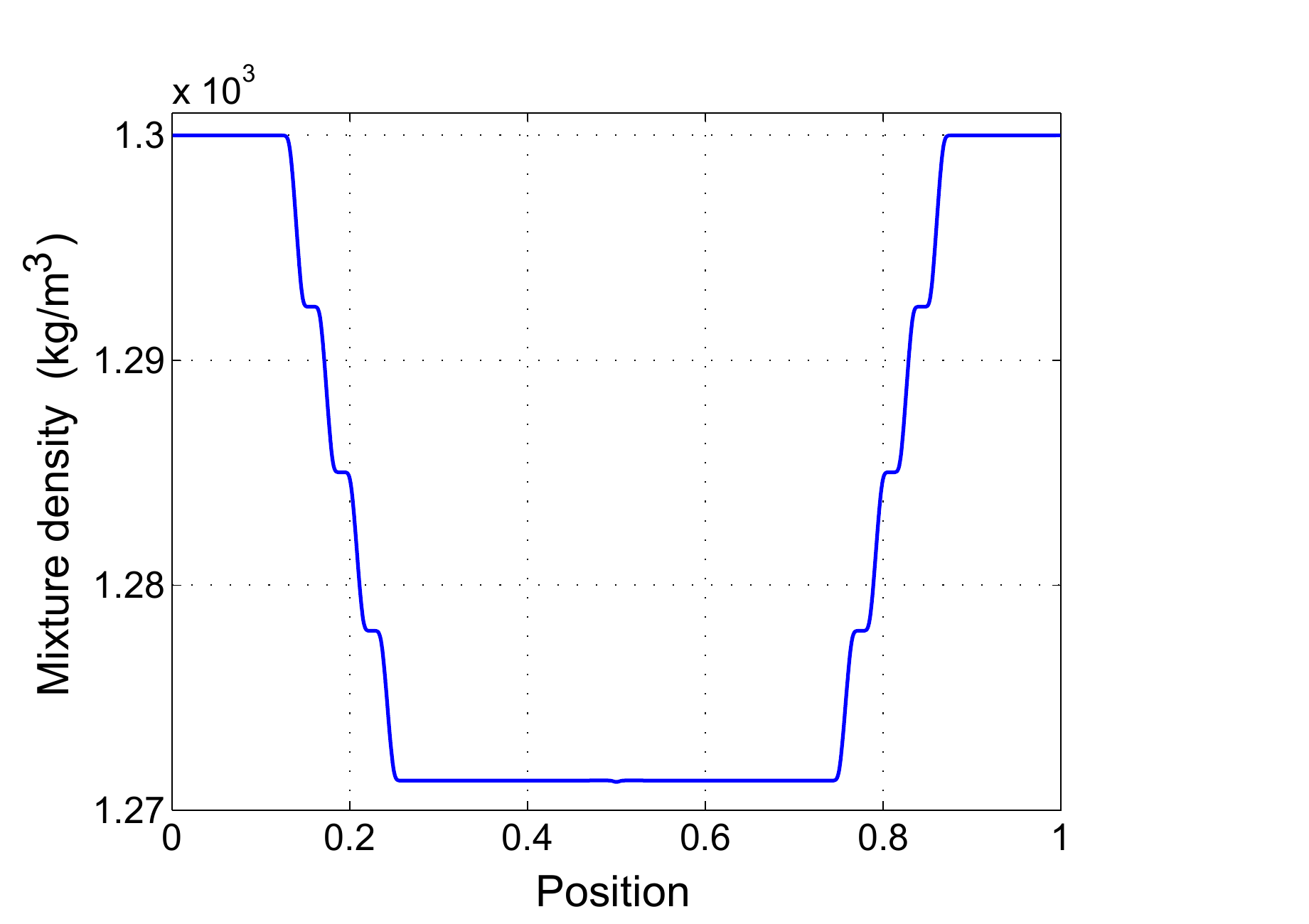}
\caption{Density profile for cavitation of the mixture of fictitious liquids at the velocity $40 m/s$}
\label{fig:expansion}
\end{figure}

The cavitation test problem has been solved for the same mixture of fictitious liquids neglecting pressure relaxation and friction and also for the isentropic case. The velocity of expansion is $U=40 m/s$, that is $u_1=u_2= u_3= u_4=-U$ on the left side and $u_1=u_2= u_3= u_4=U$ on the right hand side of the initial discontinuity. On Figure~\ref{fig:expansion}, the mixture density profile for the 3000 mesh cells at some time instant is presented. In this test problem, one can also see four left and four right propagating rarefaction waves, which is in agreement with the characteristic properties of the equations of the model.

\begin{figure}
\includegraphics*[trim = 25mm 165mm 0mm 0mm, clip, scale = 0.5]{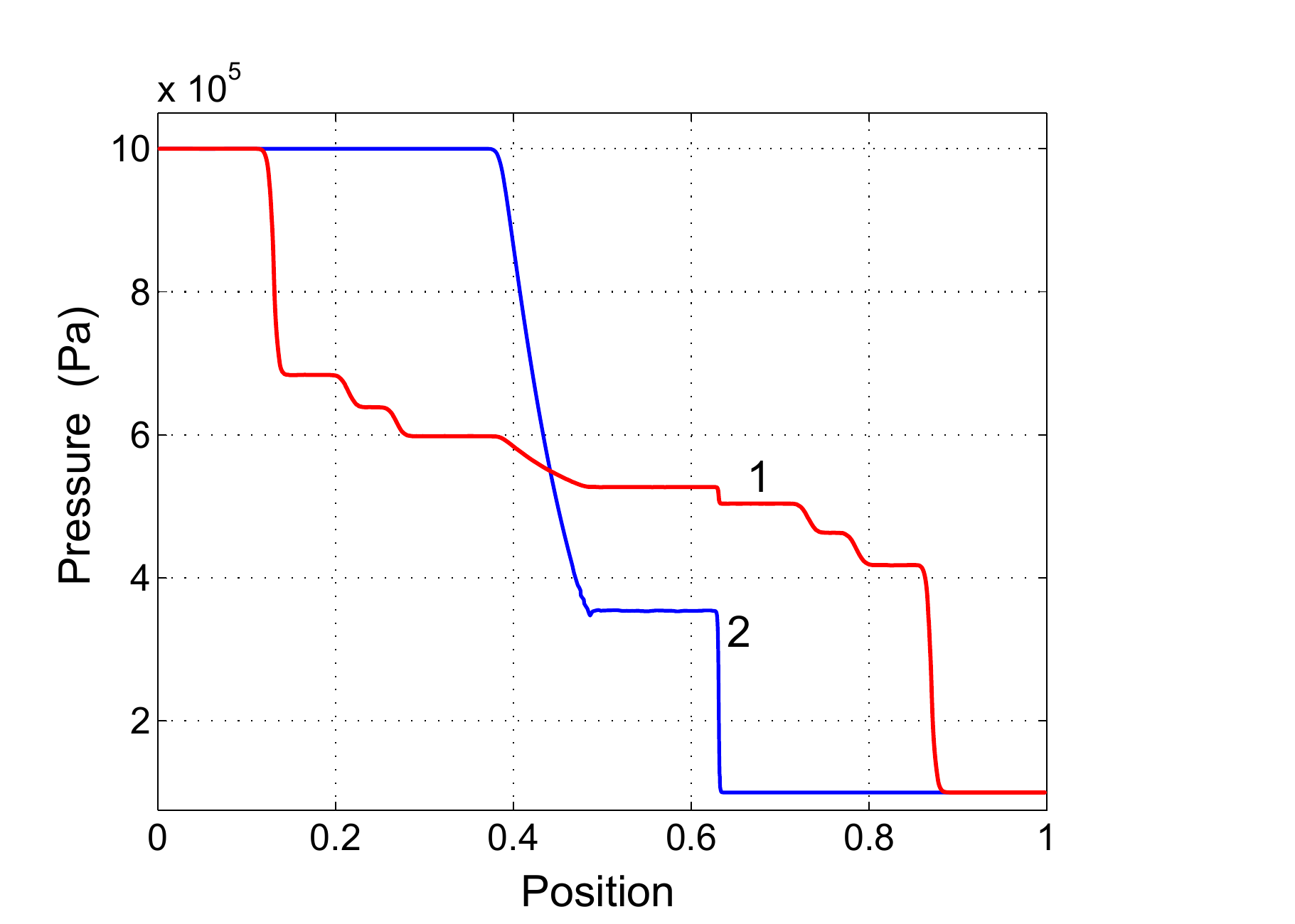}
\caption{Mixture pressure profile for the Riemann problem with the jump of pressure.
Curves 1 and 2 correspond to the computations without pressure relaxation and with instantaneous pressure relaxation accordingly.
}
\label{fig:OilMix1}
\end{figure}

For real problems, in which thermal processes and
pressure relaxation are taken into account, the wave structure can
be more complicated and sometimes it is impossible to see the wave
splitting. Here we present an example of the solution of the
Riemann problem with clearly observable splitting of shock and
rarefaction waves. Three phases are taken to be liquids with the stiffened gas EOS \eqref{stiffened_gas_EOS}
with parameters
$\rho_{01}=1600kg/m^3$, $\rho_{02}=850kg/m^3$, $\rho_{03}=1000kg/m^3$,
$\gamma_1=\gamma_2=\gamma_3=2.8$,
$C_{01}=2000m/s$, $C_{02}=1250m/s$, $C_{03}=1540m/s$,
$c_{V1}=0.96$, $c_{V2}=0.88$, $c_{V3}=4.2$. Equation of state for the
fourth phase  is an ideal gase EOS \eqref{perfect_gas_EOS} with
$\rho_{04}=0.66kg/m^3$, $\gamma_4=1.4$, $C_{04}=430m/s$, $c_{V4}=0.7$.
In the initial data for the Riemann problem velocities are zero,
the pressure on the right side is atmospheric ($10^5Pa$), and
pressure on the left side is ten times bigger ($10^6Pa$). The
phase volume fractions in the initial data are uniform:
$\alpha_1=0.7$, $\alpha_2=0.1$, $\alpha_3=0.09$, $\alpha_4=0.11$. Curves
1 and 2 on Figure~\ref{fig:OilMix1} correspond to the mixture pressure without
pressure relaxation and with instantaneous pressure relaxation
accordingly at a same moment of time. Computations have been done for 1500 mesh
cells. One can see four left propagating rarefaction waves and
four right propagating shock waves on curve 1. If the pressure relaxation is
instantaneous, then the wave structure looks like a
single right propagating shock wave and single left propagating
rarefaction wave.

\section{Conclusions}

The thermodynamically compatible system of governing equations for compressible multiphase flow is presented. The system is symmetric hyperbolic, all equations are written in a conservative form and the laws of thermodynamics hold. The choice of the equation of state in the form of the dependence of the internal energy on the parameters of state in the single entropy approximation is proposed. In order to study the properties of the model, a few Riemann test problems for the four phase flow model have been solved numerically with the finite-volume method in conjunction with the GFORCE flux. These numerical examples prove the physical reliability of the model, and hence it can be used as a theoretical basis for the study of problems of practical interest.


\bibliographystyle{amsplain}
\bibliography{bib_4phase}

\end{document}